\documentclass[letterpaper,titlepage,11pt]{article}
\pdfoutput=1
\usepackage{hyperref}

\usepackage{amssymb,amsmath,amsfonts}
\usepackage{epsfig}

\usepackage{tikz}
\usepackage{pgflibraryarrows}
\usepackage{pgflibrarysnakes}

\def\clock{{\count0=\time
          \divide\count0 60
          \ifnum\count0<10 0\fi\the\count0
          \multiply\count0 -60 \advance\count0 \time
          :\ifnum\count0<10 0\fi \the\count0
        }}
\newcommand{\timestamp}{{\small\vbox{\hbox{\tt\jobname.tex}
\hbox{\the\day/\the\month/\the\year, \clock}}}}

\newcommand{\CL}{\mathcal{L}}
\newcommand{\CO}{\mathcal{O}}
\newcommand{\CN}{\mathcal{N}}

\newcommand{\CH}{\mathcal{H}}

\newcommand{\C}{\mathbb{C}}
\newcommand{\R}{\mathbb{R}}

\newcommand{\ads}{\mbox{AdS}}

\newcommand{\nn}{\nonumber}

\newcommand{\spa}{\ , \ \ }

\newcommand{\ds}{\displaystyle}

\newcommand{\tr}{\mathop{{\rm Tr}}}

\numberwithin{equation}{section}

\begin{document}

\begin{titlepage}

\rightline{\vbox{\small\hbox{\tt NORDITA-2011-115} }}
 \vskip 2.7 cm

\centerline{\Huge \bf Holographic 3-point function at one loop}
\vskip 2 cm

\centerline{\large {\bf Agnese Bissi$\,^{1,2}$},  {\bf Troels Harmark$\,^{3}$} and
{\bf Marta Orselli$\,^{1}$} }

\vskip 1.1cm

\begin{center}
\sl $^1$ The Niels Bohr Institute, University of Copenhagen \\
\sl  Blegdamsvej 17, DK-2100 Copenhagen \O , Denmark
\vskip 0.3cm
\sl $^2$ The Niels Bohr International Academy, University of Copenhagen  \\
\sl  Blegdamsvej 17, DK-2100 Copenhagen \O , Denmark
\vskip 0.3cm
\sl $^3$ NORDITA\\
Roslagstullsbacken 23,
SE-106 91 Stockholm,
Sweden 
\end{center}
\vskip 0.7cm

\centerline{\small\tt bissi@nbi.dk, harmark@nordita.org, orselli@nbi.dk}

\vskip 1.7cm \centerline{\bf Abstract} \vskip 0.2cm \noindent
We explore the recent weak/strong coupling match of three-point functions in the AdS/CFT correspondence for two semi-classical operators and one light chiral primary operator found by Escobedo et al. This match is between the tree-level three-point function with the two semi-classical operators described by coherent states while on the string side the three-point function is found in the Frolov-Tseytlin limit. We compute the one-loop correction to the three-point function on the gauge theory side and compare this to the corresponding correction on the string theory side. We find that the corrections do not match. Finally, we discuss the possibility of further contributions on the gauge theory side that can alter our results.

\end{titlepage}

\small
\tableofcontents
\normalsize
\setcounter{page}{1}



\section{Introduction, summary and conclusion}
\label{intro}

Integrability has been the driving force behind the recent progress in the study of the spectral problem in the AdS/CFT correspondence between $\mathcal{N}=4$ Super Yang-Mills (SYM) theory and type IIB superstring theory on $\ads_5\times S^5$ (see~\cite{Beisert:2010jr} and references therein). The spectral problem consists of determining the exact spectrum of operators of the gauge theory in the planar limit and match this to the spectrum of string theory states. 

The study of the spectral problem allowed to compute the planar limit of 2-point correlation functions of gauge invariant operators from their anomalous dimension. However, to solve completely $\mathcal{N}=4$ SYM theory in the planar limit one should also know the set of all 3-point correlation functions. Thus, to have a full understanding of the AdS/CFT correspondence in the planar limit one should be able to compute the 3-point correlation functions on both the gauge theory and string theory sides, and match the two sides, possibly with the aid of integrability. However, here one faces several challenges. On the gauge theory side, it is considerably more difficult to compute 3-point functions than anomalous dimensions. In fact, even the tree-level part is highly non-trivial. On the string theory side, one needs to understand the vertex operators of string states in type IIB string theory on $AdS_5\times S^5$.

Since $\CN = 4$ SYM theory is a conformal field theory we have that 3-point correlation functions are of the form
\begin{equation}
\label{3pointfct}
\langle {\mathcal{O}}_1(x_1){\mathcal{O}}_2(x_2){\mathcal{O}}_3(x_3)\rangle = \frac{C_{123}}{|x_1-x_2|^{\Delta_1+\Delta_2-\Delta_3}|x_2-x_3|^{\Delta_2+\Delta_3-\Delta_1}|x_3 - x_1|^{\Delta_3+\Delta_1-\Delta_2}}
\end{equation}
given three gauge-invariant operators ${\mathcal{O}}_1$, ${\mathcal{O}}_2$ and ${\mathcal{O}}_3$ with definite scaling dimensions $\Delta_1$, $\Delta_2$ and $\Delta_3$. To compute the full three-point correlation function it is thus enough to compute the coefficient $C_{123}$. In the planar limit $N\rightarrow \infty$ we are only interested in the leading part of $C_{123}$ which goes like $1/N$.

Recently, progress on computing $C_{123}$ have been made on the string theory side by considering the special case of a 3-point function with two heavy (semi-classical) operators and one light chiral primary operator,
starting with the papers \cite{Zarembo:2010rr,Costa:2010rz}.%
\footnote{For recent work on holographic 3-point functions see~\cite{Roiban:2010fe,Russo:2010bt,Bissi:2011dc, Escobedo:2011xw,Georgiou:2011qk,Janik:2011bd, Kazama:2011cp, Buchbinder:2011jr}.} 
In this case, it is possible to compute the 3-point function using a prescription that employs the classical string world-sheet corresponding to the two-point function of the heavy operators. This prescription rests on the validity of the probe approximation for the supergravity state dual to the light chiral primary operator.

Building on this, a weak/strong coupling match for this type of 3-point functions was found in~\cite{Escobedo:2011xw} where the tree-level part of a 3-point function in gauge theory has been matched to the corresponding 3-point function on the string side, taking the so-called Frolov-Tseytlin limit~\cite{Frolov:2003xy, Kruczenski:2003gt}.%
 \footnote{Another interesting context in which coherent states on the gauge theory side have been compared to semi-classical string states in the AdS/CFT correspondence is in the case of non-planar corrections to folded Frolov-Tseytlin strings \cite{Casteill:2007td}.}
 The operators in the three-point function all being in the $SU(3)$ sector.%
 \footnote{Recently in \cite{Georgiou:2011qk} an analogous computation for operators in the $SL(2)$ sector was considered. Also in this case it was found perfect agreement between the weak and strong coupling result.} 
 The goal of this paper is to further explore this match by considering the one-loop correction on both sides of the correspondence.%
\footnote{Note that it was conjectured in \cite{Russo:2010bt} that both the tree-level and one-loop contributions on the gauge theory side matches the zeroth and first order contributions in the Frolov-Tseytlin limit on the string theory side for 3-point function of the kind we are examining in this paper.}

The Frolov-Tseytlin limit was originally conceived as a limit of classical string solutions of the bosonic sigma-model on $\ads_5 \times S^5$. In the case of a string moving on $S^5$ with angular momentum $J$, the energy of the string is expanded in a limit of large $J$ around a BPS solution with the expansion parameter $\lambda / J^2$. This expansion can then be compared to the loop expansion on the gauge theory side. The expansion coefficients match the gauge theory side up to and including the second order in the expansion parameter, meaning two-loops on the gauge theory side, but the matching breaks down at three-loops~\cite{Callan:2003xr,  Callan:2004uv}. In \cite{Harmark:2008gm} it is shown that the match at one-loop is not a coincidence but instead a result of the quantum corrections to the string being suppressed near the BPS point, enabling one to consider a regime where the classical action of the string is large even if one approaches weak 't Hooft coupling.

The understanding of the Frolov-Tseytlin limit was further enhanced with the work of Kruczenski \cite{Kruczenski:2003gt}. There it is shown how for semi-classical operators on the gauge theory side one can use a coherent state description thus enabling one to write down an effective sigma-model description. Hence, one can directly compare the sigma-model action for semi-classical operators on the gauge theory side to the classical sigma-model action on the string theory side in the Frolov-Tseytlin limit. In particular, if we consider operators in the $SU(3)$ sector we find on both the gauge theory and string theory sides the energy (scaling dimension)
\begin{equation}
\label{Eexpan}
E = J + \frac{\lambda}{2J} \int_0^{2\pi} \frac{d\sigma}{2\pi} \partial_\sigma {\bf \bar{u}} \cdot \partial_\sigma {\bf u} + \CO( \lambda^2 / J^3 )
\end{equation}
with the non-linear sigma-model field ${\bf u}(\tau,\sigma)$ taking values in $\C^3$ and being a solution of the equations of motion (EOMs) following from using \eqref{Eexpan} as the Hamiltonian supplemented by the constraint ${\bf u} \cdot \partial_\sigma \bar{{\bf u}} = 0$.

The work of~\cite{Escobedo:2011xw} can thus be seen as a natural extension of the work of \cite{Frolov:2003xy, Kruczenski:2003gt}  to 3-point correlation functions, using the prescription of \cite{Zarembo:2010rr} for two semi-classical operators and a light chiral primary operator in the $SU(3)$ sector of $\CN=4$ SYM theory. Amazingly, they found on both the gauge theory and the string theory side the same result
\begin{equation}
\label{escobedoC123}
C^{(0)}_{123} = \frac{J}{N}  \frac{(j_2+j_3)! }{j_2! j_3!} \sqrt{ \frac{j_1! j_2! j_3! }{(j_1+j_2+j_3-1)!} } \int_0^{2\pi} \frac{d\sigma}{2\pi} \bar{u}_1^{j_1} u_2^{j_2} u_3^{j_3} 
\end{equation}
The gauge theory operators are constructed from the three complex scalars $Z$, $X$ and $Y$ of $\CN=4$ SYM theory and their complex conjugates $\bar{Z}$, $\bar{X}$ and $\bar{Y}$. The $\CO_1$ operator is made of $J_1+j_1$ $\bar{Z}$'s, $J_2-j_2$ $\bar{X}$'s and $J_3-j_3$ $\bar{Y}$'s, the $\CO_2$ operator of $J_1$ $Z$'s, $J_2$ $X$'s and $J_3$ $Y$'s and the $\CO_3$ operator of $j_1$ $Z$'s, $j_2$ $\bar{X}$'s and $j_3$ $\bar{Y}$'s. We introduce the quantity $J=J_1+J_2+J_3$. Note that, by construction, this is a non-extremal 3-point function for $j_2+j_3\neq 0$. 
While $\CO_3$ is taken to be a $1/2$ BPS chiral primary operator, $\CO_1$ and $\CO_2$ are constructed as coherent states with corresponding sigma-model fields ${\bf u}(\tau,\sigma)$ and $\bar{{\bf u}}(\tau,\sigma)$, respectively. Here ${\bf u}=(u_1,u_2,u_3)$ is a solution of the EOMs following from the one-loop Hamiltonian \eqref{Eexpan}. The coefficient \eqref{escobedoC123} is then computed at tree-level by doing Wick contractions. 
On the string theory side, one considers the leading part of $C_{123}$ in the Frolov-Tseytlin limit of the corresponding 3-point function using the prescription of \cite{Zarembo:2010rr}.

In this paper we explore whether the match of the 3-point correlation function coefficient \eqref{escobedoC123}  between the gauge theory and string theory sides can be extended beyond tree-level on the gauge theory side to include the one-loop correction, corresponding to the first order in the Frolov-Tseytlin expansion parameter $\lambda / J^2$ on the string theory side.

On the gauge theory side we write the tree-level and one-loop part as
\begin{equation}
\label{onel}
C_{123} = C^{(0)}_{123} + \lambda C^{(1)}_{123} + \cdots
\end{equation}
where $C_{123}$ is the coefficient in \eqref{3pointfct}. Note that in \eqref{3pointfct} we use the renormalized operators thus $C^{(1)}_{123}$ is the scheme-independent part of the one-loop coefficient. To simplify our computation we consider a special class of operators with $J_3=j_3=0$ and $j_1=j_2=j$. Then all three operators are in an $SU(2)$ sector of $\CN=4$ SYM theory (note that obviously they are not in the same $SU(2)$ sector). The leading order part \eqref{escobedoC123} takes the form
\begin{equation}
\label{escobedoC123b}
C^{(0)}_{123} = \frac{1}{N}  \frac{  j! J }{ \sqrt{ (2j-1)! } } \int_0^{2\pi} \frac{d\sigma}{2\pi} ( \bar{u}^1 u_2 )^{j} 
\end{equation}
For this class of operators we compute  the one-loop correction $C^{(1)}_{123}$ to the 3-point correlation function coefficient on both the gauge theory and the string theory side.

On the gauge theory side the one-loop correction diagrams contributing to $C^{(1)}_{123}$ can be computed in the planar limit using the prescriptions given in \cite{Okuyama:2004bd,Roiban:2004va,Alday:2005nd}. However, there is another contribution as well~\cite{Beisert:2002bb, Kristjansen:2010kg, Grossardt:2010xq}. The origin of this is that our computation should be thought of as the first correction in an all-order series in powers of $\lambda$. Thus, since ${\bf u}(\tau,\sigma)$ receives corrections at order $\lambda$ from considering the two-loop contribution to the effective sigma-model description \eqref{Eexpan}, these corrections also contribute to the 3-point function. Writing ${\bf u} = {\bf u}^{(0)} + \lambda J^{-2} {\bf u}^{(1)} + \CO( \lambda^2 )$ we can find this type of corrections simply by substituting in the full ${\bf u}$ in $C^{(0)}_{123}$ of Eq.~\eqref{escobedoC123b} and extracting the $\lambda$ corrections. Combining both contributions to the one-loop correction, we find
\begin{eqnarray}
\label{gaugesideC123}
C_{123} &=& \frac{1}{N}  \frac{  j! J }{ \sqrt{ (2j-1)! } } \int_0^{2\pi} \frac{d\sigma}{2\pi} ( \bar{u}^1 u_2 )^{j}  \left[ 1-\frac{\lambda}{2J^2} \Big\{ \partial_{\sigma}\bar{\bf u}\cdot\partial_{\sigma}{\bf u} \right.  \nn \\[4mm] && \left. \left. + \frac{j^2-1}{2} \Big( \frac{\partial_\sigma(\bar{u}^1 u_2)}{\bar{u}^1 u_2}  \Big)^2+  \frac{\partial_\sigma\bar{u}^1 \partial_\sigma u_2}{\bar{u}_1 u_2} 
\right\}
\right] 
+ \CO (\lambda^2)
\end{eqnarray}
where ${\bf u} = {\bf u}^{(0)} + \lambda J^{-2} {\bf u}^{(1)} + \CO( \lambda^2 )$ is the full sigma-model field including the correction coming from the two-loop Hamiltonian. The full computation leading to Eq.~\eqref{gaugesideC123} is laid out in detail in Section \ref{gauge} of this paper. Moreover, in Section \ref{sec:spinflip} we discuss the possibility of a further contribution to the result \eqref{gaugesideC123} coming from the so-called spin-flipped coherent state \cite{Kruczenski:2004kw}.

Turning to the string theory side the prescription of computing the 3-point correlation function coefficient $C_{123}$ for two semi-classical operators and a light chiral primary operator is \cite{Zarembo:2010rr}
\begin{equation}
\label{zaremboC123}
C_{123} = c_j \frac{\sqrt{\lambda}}{N}\int_{-\infty}^{+\infty} d{\tau_e} \int_{0}^{2 \pi} \frac{d\sigma}{2\pi}\frac{ (\bar{U}^1 U_2 )^{j}  }{\cosh^{2j} (\frac{\tau_e}{\kappa})} \left[ \frac{2}{\kappa^2\cosh^2  (\frac{\tau_e}{\kappa})}-\frac{1}{\kappa^2}-\partial_a {\bf \bar U}\cdot\partial^a {\bf U}\right]
\end{equation}
for the class of operators we are considering, with $c_j$ a function only of $j$ and $\tau_e$ is the Euclidean time. Here ${\bf U}(\tau,\sigma)$ takes values on $\C^3$ and describes the embedding of the type IIB string on $S^5$. In the Frolov-Tseytlin limit we consider the fluctuations around a point particle moving with angular momentum $J$ around one of the equators. We can write ${\bf U} = e^{i\tau/\kappa} {\bf u}$ where ${\bf u}$ describes the fluctuations and $\kappa = \sqrt{\lambda}/J $.~\footnote{Note that $\kappa$ in this paper corresponds to $1/\kappa$ used in \cite{Escobedo:2011xw}.} The Frolov-Tseytlin limit in our notation is then $\kappa \rightarrow 0$ with $\frac{1}{\kappa} \partial_\tau {\bf u}$ and $\partial_\sigma {\bf u}$ fixed. This gives an expansion in $\kappa^2 = \lambda / J^2$ which parallels the loop expansion on the gauge theory side. 

Performing now the Frolov-Tseytlin expansion in \eqref{zaremboC123}, as well as in the bosonic string sigma-model on $\R \times S^5$,  we find
\begin{equation}
\label{stringsideC123}
C_{123} = \frac{1}{N}  \frac{  j! J }{ \sqrt{ (2j-1)! } } \int_0^{2\pi} \frac{d\sigma}{2\pi} ( \bar{u}^1 u_2 )^{j} \left( 1- \frac{\lambda}{J^2} \frac{(2j+1)}{2j}\partial_{\sigma}\bar{\bf u}\cdot\partial_{\sigma}{\bf u} \right)+ \CO (\lambda^2/J^4)
\end{equation}
It is important to remark here that ${\bf u}$ obeys the same EOMs as those following from the two-loop extension of the Hamiltonian \eqref{Eexpan}. This is due to the well-known fact, shown explicitly in \cite{Frolov:2003xy, Kruczenski:2003gt, Kruczenski:2004kw, Minahan:2005mx, Minahan:2005qj}, that the sigma-models in the Frolov-Tseytlin limit following Kruczenski's work match for both one- and two-loops. Therefore, ${\bf u}$ in \eqref{stringsideC123} as well as in \eqref{gaugesideC123} obeys the same corrected EOMs, $i.e.$ it includes the same $\lambda/J^2$ corrections in ${\bf u} = {\bf u}^{(0)} + \lambda J^{-2} {\bf u}^{(1)} + \CO( \lambda^2/J^4 )$ on both sides of the correspondence. This means that it is more convenient not to write these corrections to $C_{123}$ explicitly as they already are guaranteed to match from the fact that the $C^{(0)}_{123}$ in \eqref{escobedoC123b} matches. The full computation of \eqref{stringsideC123} is described in detail in Section \ref{string}.

Comparing now $C_{123}$ on the gauge theory side \eqref{gaugesideC123} and the string theory side \eqref{stringsideC123} we see that they do not match. 
This means that the match between the tree-level gauge theory answer and the leading Frolov-Tseytlin limit on the string side \eqref{escobedoC123b} found in~\cite{Escobedo:2011xw} does not extend beyond one-loop. However, we discuss below the possibility of further contributions on the gauge theory side to the 3-point coefficient Eq.~\eqref{gaugesideC123}.

An immediate question to ask now is whether one should have expected a match, or in other words, whether a mismatch is consistent with our current knowledge and hypothesis surrounding the AdS/CFT correspondence. The answer is that a mismatch does not contradict anything of what we know about the AdS/CFT correspondence. First of all, the result \eqref{gaugesideC123} is computed at weak 't Hooft coupling $\lambda \ll 1$ and \eqref{stringsideC123} at strong 't Hooft coupling $\lambda \gg 1$. Thus, without further arguments, it is by no means clear why there should be any match at all, even for the leading result \eqref{escobedoC123b}. 

The only known possible line of argument for a match of the two sides seems to be the one of~\cite{Harmark:2008gm}. In \cite{Harmark:2008gm} it is argued why the one-loop part of the energy/scaling dimension in \eqref{Eexpan} matches. The central point of the argument is to consider perturbations around a protected BPS state with energy $J$, the leading contribution in~\eqref{Eexpan}. This means that one can consider a regime with $E - J \ll \lambda \ll 1$ and $J \gg 1$. In~\cite{Harmark:2008gm} it is then shown that it is possible to take a limit on the string side that zooms in to this regime, which requires taking $\lambda$ to go to zero. Said briefly, this is possible because the effective string tension in front of the sigma-model action in this limit becomes $J$ rather than $\sqrt{\lambda}$ and because the quantum corrections, either from the background, or from the fields that decouple in the limit, are suppressed.

However, unfortunately it seems that the line of arguments of \cite{Harmark:2008gm} does not extend to the type of 3-point functions considered above. This is because the tree-level part ($i.e.$ leading part of Frolov-Tseytlin limit) of $C_{123}$ is not a protected quantity. Instead the tree-level answer \eqref{escobedoC123b} receives corrections in powers of $1/J$ as compared to the corresponding 3-point coefficient $C^{BPS}_{123}$ for chiral primary operators with the same charges. Considering now $C_{123} / C^{BPS}_{123} - 1$ the leading correction on the gauge theory side comes from the tree-level part and goes like $1/J^2$. Instead on the string theory side we have $\lambda / J^2$ corrections coming from the first order part of the Frolov-Tseytlin expansion in $\lambda/J^2$. And since $\lambda \gg 1$ these dominate over $1/J^2$, unlike on the gauge theory side. 

While no known argument exists for a match of $C_{123}$ for the class of 3-point functions considered here, it is still worth considering whether the mismatch that we find could stem from overlooked subtleties on either the gauge theory or the string theory side. 

On the gauge theory side we use the prescription of \cite{Okuyama:2004bd,Roiban:2004va,Alday:2005nd}. This seems a physically sound prescription, as it consists in computing all the 3-point diagrams involving all three operators using the one-loop Hamiltonian and summing them up. However, it would be prudent to validate further this prescription by making checks for explicit examples, such as in \cite{Grossardt:2010xq}. On the string theory side we use the prescription of \cite{Zarembo:2010rr} which is equivalent to that of \cite{Costa:2010rz} for computing 3-point correlation functions of two semi-classical operators and one light chiral primary operator. But it is still unclear whether this is the right prescription for this computation~\cite{Janik:2011bd, Kazama:2011cp, Buchbinder:2011jr}.
Note that 
 on the gauge theory side, one needs to address the subtlety that the two semi-classical states which are approximated with the same coherent state have to be slightly different due to conservation of the R charges. In \cite{Escobedo:2011xw} it is argued using numerics that the right prescription is that the two operators differ by a zero mode. However, it is not clear that this extends to one-loop, thus it would be useful if one could numerically test our gauge theory result \eqref{gaugesideC123} for explicit examples of operators. Finally, as already mentioned above, in Section \ref{sec:spinflip} we discuss the possibility of a further contribution to the result \eqref{gaugesideC123} coming from the so-called spin-flipped coherent state \cite{Kruczenski:2004kw}.

In conclusion, we think that the study of the 3-point correlation functions in the planar limit of the AdS/CFT correspondence is a highly fascinating new avenue to follow and that it would be very interesting if the techniques of integrability could be extended to this as well. With this in hand, one could possibly understand how the 3-point coefficients \eqref{gaugesideC123} and \eqref{stringsideC123} can interpolate from weak to strong coupling. This is clearly an interesting problem that deserves further investigation also in view of the fact that a similar comparison between the weak and the strong coupling result in the case of 2-point correlation functions was crucial in establishing a connection between the two opposite regimes. A similar study in the case of 3-point correlation functions would be important in deriving an all loop result.



\section{Gauge theory side: Non-extremal 3-point functions}
\label{gauge}

In this section we compute the one-loop correction to the planar limit of a non-extremal 3-point function with two heavy (semi-classical) operators and one light chiral primary operator, each in a separate $SU(2)$ sector of $\CN=4$ SYM. Following \cite{Escobedo:2011xw} we use coherent states to approximate the two heavy operators. The one-loop correction is computed using methods of \cite{Okuyama:2004bd,Roiban:2004va,Alday:2005nd}. The result of the computation is the formula \eqref{gaugesideC123} listed in the Introduction. 

Below we describe the gauge theory operators, setup the notation for our computation and briefly review the sigma-model description of the two heavy operators. In Section \ref{sec:treelevel} we review the tree-level part of the result found in \cite{Escobedo:2011xw}. In Section \ref{sec:oneloop} we compute the one-loop correction to the tree-level result including both the contribution coming from requiring the two heavy operators to correspond to eigenstates of the two-loop correction of the dilation operator, and the contribution coming from one-loop diagrams for the 3-point function. Finally, in Section \ref{sec:spinflip}, we consider the possible contribution due to the correction to the coherent state description from spin-flipped coherent states. 

\subsubsection*{The three operators}

The three operators $\CO_i (x_i)$, $i=1,2,3$, for which we compute the 3-point function \eqref{3pointfct} are given as follows. All three operators are in the scalar sector of $\mathcal{N}=4$ SYM theory and we consider single trace operators made out of three complex scalars $Z$, $X$ and $Y$. Moreover, each operator is in an $SU(2)$ sector of $\CN=4$ SYM. Specifically, $\CO_1(x_1)$ is made of $J_1+j$ $\bar Z$ scalars and $J_2-j$ $\bar X$ scalars and has length $J = J_1 + J_2$. $\CO_2(x_2)$ is made of $J_1$ ${Z}$ scalars and $J_2$ ${X}$ scalars and also has length $J$. $\CO_3(x_3)$ is the $1/2$ BPS chiral primary operators made of $j$ $Z$ scalars and $j$ $\bar{X}$ scalars.
Note that this gives a non-extremal 3-point function for any non-zero $j$.

In more detail, the $\CO_1(x_1)$ and $\CO_2(x_2)$ operators are semi-classical operators thus with $J \gg 1$, and are written as 
\begin{equation}
\label{CO1}
\CO_{1} (x_1) = \CN_1  \bar{{\bf u}}^{i_1} ( \frac{k+1}{l} ) \,  \bar{{\bf u}}^{i_2} ( \frac{k+2}{l} ) \cdots  \bar{{\bf u}}^{i_{J}} (\frac{k}{l}) : \tr ( \bar{W}_{i_1} \bar{W}_{i_2} \cdots \bar{W}_{i_{J}} ) : (x_1)
\end{equation}
\begin{equation}
\label{CO2}
\CO_{2} (x_2) = \CN_2 {{\bf v}}_{j_1} (\frac{k+1}{l} ) \, {{\bf v}}_{j_2}(\frac{k+2}{l} ) \cdots {{\bf v}}_{j_{J}} ( \frac{k}{l}) : \tr ( {W}^{j_1} {W}^{j_2} \cdots {W}^{j_{J}} ) : (x_2)
\end{equation}
with
\begin{equation}
W^i = (Z,X) \spa \bar{W}_i = (\bar{Z},\bar{X})
 \spa l \equiv \frac{J}{2\pi}
\end{equation}
Here ${{\bf u}} (\sigma)$ and ${{\bf v}} (\sigma)$ correspond for each site of the single trace operators to coherent states in the spin $1/2$ representation of $SU(2)$. Specifically the $k$'th site is at $\sigma= k /l$ and the functions ${{\bf u}} (\sigma)$ and ${{\bf v}} (\sigma)$ are periodic in $\sigma$ with period $2\pi$ and they take values in $\C^2$. That the two operators are semi-classical also means that the functions ${{\bf u}} (\sigma)$ and ${{\bf v}} (\sigma)$ are slowly varying in $\sigma$. 
The third operator $\CO_3(x_3)$ we can write as
\begin{equation}
\CO_{3} (x_3) = \CN_3  : \tr ( \mbox{sym} (\bar{X}^j Z^{j} ) ) : (x_3)
\end{equation}
It is important to note that we did not include the corrections to the coherent state description of the operators \eqref{CO1}-\eqref{CO2} from the so-called spin-flipped coherent states \cite{Kruczenski:2004kw}. We consider the effect of this in Section \ref{sec:spinflip}.

We want to compute the one-loop correction to the coefficient $C_{123}$ appearing in Eq.~\eqref{3pointfct}.
We therefore write 
\begin{equation}
C_{123}=C_{123}^{(0)}+\lambda' C_{123}^{(1)}+\CO(\lambda'^2)
\label{c123new}
\end{equation}
where we introduced the quantity $\lambda'=\lambda/J^2$ since this is the parameter that naturally appears in the expansion.

In the following we present a brief summary of the Landau-Lifshitz model for semiclassical operators,
then we review the computation of the leading term $C_{123}^{(0)}$ and explain how we compute the one-loop, scheme-independent, contribution $C_{123}^{(1)}$.


\subsubsection*{The sigma-model description}

Semiclassical operators in the $SU(2)$ sector of $\CN=4$ SYM theory can be described using the Landau-Lifshitz sigma model \cite{Kruczenski:2003gt}. 

In the planar limit the one-loop correction to the dilatation operator of $\CN=4$ SYM theory can be regarded as the Hamiltonian
\begin{equation}
H=\frac{\lambda}{8\pi^2}\sum_{l=1}^J\left(I_{l,l+1}-P_{l,l+1}\right)
\label{d2}
\end{equation}
where $I_{l,l+1}$ is the identity operator and $P_{l,l+1}$ is the permutation operator.
Acting with this Hamiltonian on a semiclassical operator such as $\CO_1$ one can compute the energy which, up to one loop order, is given by
\begin{equation}
\label{llenergy}
E\simeq J\left(1+\frac{\lambda'}{2 }\int_0^{2\pi} \frac{d\sigma}{2\pi} \partial_\sigma {\bf \bar{u}} \cdot \partial_\sigma {\bf u} \right)
\end{equation}
for large $J$, where $J$ is the bare scaling dimension of the operator. 

For use below we note that it was found in \cite{Escobedo:2011xw} that one can choose a gauge for the local $U(1)$ phase symmetry transformation of ${\bf u} (\sigma) \rightarrow e^{i \Lambda (\sigma)} {\bf u} (\sigma)$ such that 
\begin{equation}
\label{gaugevir}
{\bf \bar u} ' \cdot {\bf u } = {\bf \bar u}  \cdot {\bf u }' = 0
\end{equation}
This is in accordance with the Virasoro constraint on the string theory side.

As anticipated in Section~\ref{intro}, the two-loop contribution to the Landau-Lifshitz sigma model generates corrections of order $\lambda$ to the coherent state function ${\bf u}(\sigma)$ that describes the gauge theory operators. Since these corrections also contribute to the computation of the 3-point function coefficient $C_{123}$ at one-loop, we have to take them into account. The wave function ${\bf u}(\sigma)$ appearing in our expressions should therefore be solution of the EOMs of the Landau-Lifshitz model up to two-loops.

To write down the two-loop contribution to the sigma model, it is convenient to use the following notation. We are considering gauge theory operators in the $SU(2)$ sector of $\CN=4$ SYM on ${\mathcal R} \times S^3$. 
To obtain a sigma-model description of single trace operators we introduce a coherent state $| \vec{n} \rangle$ for each site of the trace such that
\begin{equation}
\langle \vec{n} | \vec{\sigma} | \vec{n} \rangle = \vec{n}
\end{equation}
where $\vec{\sigma}$ are the two by two Pauli matrices and
$\vec{n}$ is a unit vector pointing to a point on the two-sphere
parameterized as
\begin{equation}
\label{vecn} \vec{n} = (\cos \theta \cos \varphi, \cos \theta \sin
\varphi, \sin \theta )
\end{equation}
In the limit $J\rightarrow \infty$ the Lagrangian of the Landau-Lifshitz model up to two loops reads~\cite{Kruczenski:2004kw}
\begin{equation}
\label{LLlagrg}
 \CL_{LL}= \frac{1}{2} \sin \theta \dot{\varphi} - \frac{\lambda'}{8} (\vec{n}')^2 +  \frac{\lambda'^2}{32} \Big[ (\vec{n}'')^2 - \frac{3}{4} (\vec{n}')^4 \Big] + \CO (\lambda'^3 )
\end{equation}
where prime denotes derivatives with respect to the continuos variable $\sigma$ which can be introduced to describe the trace in the limit $J \rightarrow \infty$.
$\sigma$ is periodic with period $2\pi$ therefore we 
map the $k$'th site to $\sigma=2\pi k/J$ and we consider the field $\vec{n} ({t}, \sigma)$. 
Accordingly the discrete sum over the sites of the single trace operators is mapped to the integral
$\frac{J}{2\pi} \int_{0}^{2\pi} d\sigma$. Moreover, in deriving \eqref{LLlagrg} one also uses that
\begin{equation}
\vec{n}_{k+1} - \vec{n}_{k} = \exp \left( \frac{2\pi}{J}
\partial_\sigma \right) \vec{n} - \vec{n}
\end{equation}
It is important to note that the two-loop Lagrangian \eqref{LLlagrg} is derived by including the effect of spin-flipped coherent state \cite{Kruczenski:2004kw}. We discuss the correction from spin-flipped coherent states to the heavy operators \eqref{CO1}-\eqref{CO2} in Section \ref{sec:spinflip}.

Instead of using the unit vector ${\vec n}$, in the rest of this section we will describe the semiclassical operators $\CO_1$ and $\CO_2$ using the complex functions 
${\bf u}(\sigma)$ and ${\bf v}(\sigma)$ obeying the condition ${\bf \bar u}\cdot {\bf u}=1$ and ${\bf \bar v}\cdot {\bf v}=1$.


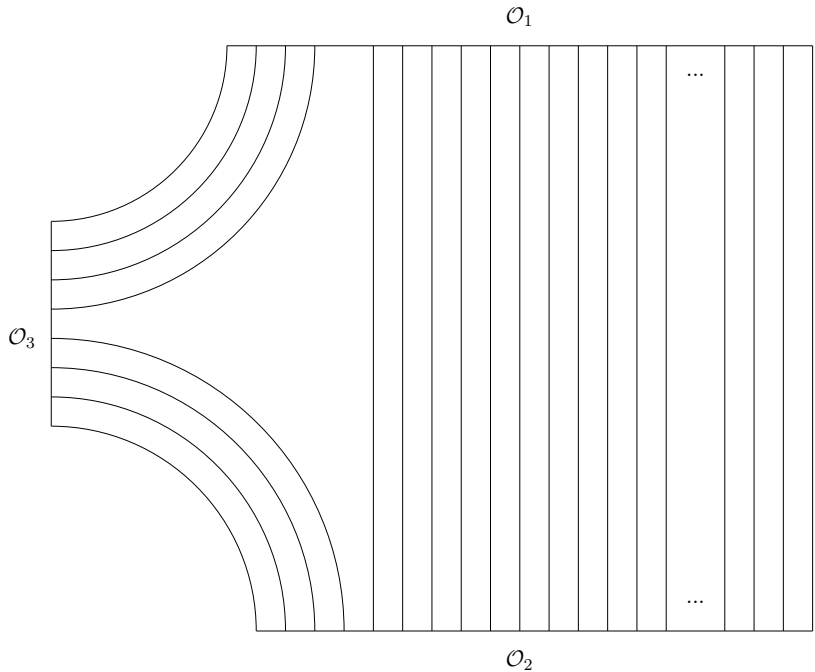
\begin{figure}
\begin{center}
\resizebox{11cm}{!}{
\begin{tikzpicture}
\draw (0.5,0) to (10,0);
\draw (0,10) to (10,10);
\draw (-3,3.5) to (-3,7);
\draw[](-3,7)arc(270:360:3cm);
\draw[](0.5,0)arc(0:90:3.5cm);
\draw[](1,0)arc(0:90:4cm);
\draw[](1.5,0)arc(0:90:4.5cm);
\draw[](2,0)arc(0:90:5cm);
\draw[](-3,6.5)arc(270:360:3.5cm);
\draw[](-3,6)arc(270:360:4cm);
\draw[](-3,5.5)arc(270:360:4.5cm);
\draw (8,0.5) node(x)  {...};
\draw (8,9.5) node(x)  {...};
\draw (2.5,0) to (2.5,10);
\draw (3,0) to (3,10);
\draw (3.5,0) to (3.5,10);
\draw (4,0) to (4,10);
\draw (4.5,0) to (4.5,10);
\draw (5,0) to (5,10);
\draw (5.5,0) to (5.5,10);
\draw (6,0) to (6,10);
\draw (6.5,0) to (6.5,10);
\draw (7,0) to (7,10);
\draw (7.5,0) to (7.5,10);
\draw (8.5,0) to (8.5,10);
\draw (9,0) to (9,10);
\draw (9.5,0) to (9.5,10);
\draw (10,0) to (10,10);
\draw (5,-0.5) node(x)  {$\mathcal{O}_2$};
\draw (5,10.5) node(x)  {$\mathcal{O}_1$};
\draw (-3.5,5) node(x)  {$\mathcal{O}_3$};
\end{tikzpicture}
}
\end{center}
\caption{\label{fig:tree} Tree level contractions between $\CO_1$, $\CO_2$ and $\CO_3$.}
\end{figure}


\subsection{Tree-level computation}
\label{sec:treelevel}

We now review the computation of the leading planar contribution to $\langle {\CO}_{1} (x_1) \CO_{2} (x_2) \CO_{3} (x_3) \rangle$ at tree-level~\cite{Escobedo:2011xw}.
Define 
\begin{equation}
\label{b}
B \equiv \prod_{m=1}^{J}  \bar{{\bf u}} ( \frac{m}{l} ) \cdot {\bf v} ( \frac{m}{l} )
\end{equation}
Note that $B$ depends on ${\bf u}(\sigma)$ and ${\bf v}(\sigma)$ but not on the choice of $k$. Our convention for the tree-level 3-point diagram is that we contract the $j$ first letters of ${\CO}_{1}$ with $\CO_{3}$ and the rest is then contracted with $\CO_{2}$. Also, we contract the $j$ first letters of ${\CO}_{2}$ with $\CO_{3}$ and the rest with ${\CO}_{1}$ (see Fig.~\ref{fig:tree}).
Disregarding propagators, combinatoric factors and such, the tree-level contractions give
\begin{equation}
A(k) = B \prod_{m=k+1}^{k+j} \frac{\bar{{u}}^1 ( \frac{m}{l} ) {v}_2 (\frac{m}{l} )}{\bar{{\bf u}} (\frac{m}{l} ) \cdot {\bf v} (\frac{m}{l} )}
\end{equation}
Including the sum over $k$, we have
\begin{equation}
\sum_{k} A(k) = B \sum_k \prod_{m=k+1}^{k+j} \frac{\bar{u}^1 (\frac{m}{l} ) v_2 ( \frac{m}{l} )}{\bar{{\bf u}} (\frac{m}{l} ) \cdot {\bf v} (\frac{m}{l} )}
\end{equation}
Since $\bar{{\bf u}}$ varies slowly and $j \ll J$, the difference for $\bar{{\bf u}}$ at two different values of $\sigma$ can be estimated using a Taylor expansion. Similarly can be done for ${\bf v}$. We find
\begin{equation}
\sum_{k} A(k) = B \sum_k \left(  \frac{(\bar{u}^1  v_2 ) (\frac{k}{l} )}{(\bar{{\bf u}}  \cdot {\bf v} ) (\frac{k}{l} )} \right)^j \left( 1 +\frac{j(j+1)}{2\, l} \left(  \frac{(\bar{u}^1 v_2)' (\frac{k}{l} ) }{(\bar{u}^1 v_2 ) (\frac{k}{l} )} -\frac {( \bar{{\bf u}} \cdot {\bf v}  )'(\frac{k}{l} )}{(\bar{\bf u} \cdot {\bf v}) ( \frac{k}{l} )}\right) + \cdots \right)
\end{equation}
with prime denoting the derivative with respect to $\sigma$.

Thus, approximately we find
\begin{equation}
\sum_{k} A(k) \simeq B \sum_k \left(  \frac{(\bar{u}^1  v_2 ) (\frac{k}{l} )}{(\bar{{\bf u}}  \cdot {\bf v} ) ( \frac{k}{l} )} \right)^j \simeq B\, J \int_0^{2\pi} \frac{d\sigma}{2\pi} \left(  \frac{(\bar{u}^1  v_2 ) ( \sigma )}{(\bar{{\bf u}}  \cdot {\bf v} ) ( \sigma )} \right)^j
\label{gtleading}
\end{equation}
We now want to use the approximation ${\bf v} (\sigma) = {\bf u} (\sigma)$ in \eqref{gtleading}.
Indeed, naively one can say that since $\CO_3$ is a small operator, $\CO_2$ is to a good approximation the complex conjugate of $\CO_1$ for $j \ll J$. However, as explained in \cite{Escobedo:2011xw}, to make sure that the difference between $\CO_1$ and $\CO_2$ does not enter in the result \eqref{gtleading} to leading order in a $j/J$ expansion one needs to specify how the difference between $\CO_1$ and $\CO_2$ in detail is realized. It is found in \cite{Escobedo:2011xw} that constructing $\CO_2$ from $\CO_1$ by adding more roots to already existing classical cuts one ensures that the difference ${\bf v} (\sigma) - {\bf u} (\sigma)$ is of order $j/J$ which is enough to guarantee that the difference between ${\bf u}$ and ${\bf v}$ does not enter to leading order. For use later below, we parameterize the difference between $\CO_1$ and $\CO_2$ as
\begin{equation}
\label{vu_app}
{\bf v} (\sigma) = {\bf u} (\sigma) + \frac{j}{l} {\bf \delta u} (\sigma)
\end{equation}
Using now \eqref{vu_app} in \eqref{gtleading} we find that $B=1$ to leading order in $j/J$ and hence
\begin{equation}
\label{newescobedoC123b}
C^{(0)}_{123} = \frac{\CN_3}{N} \sum_k A(k)= \frac{1}{N}  \frac{  j! J }{ \sqrt{ (2j-1)! } } \int_0^{2\pi} \frac{d\sigma}{2\pi} ( \bar{u}^1 u_2 )^{j} 
\end{equation}
up to finite size corrections in $1/J$, where we used that $\CN_3=\frac{j!}{\sqrt{(2j-1)!}}$.
This is obtained using that among all possible terms in $\CO_3$ only $\tr\left({\bar X}^j Z^j\right)$ can give a non-zero contribution to the Wick contractions. This is the same result given in Eq.~\eqref{escobedoC123b} and already derived in \cite{Escobedo:2011xw}. In the following we extend this result to include the one-loop correction.


\subsection{One-loop computation}
\label{sec:oneloop}

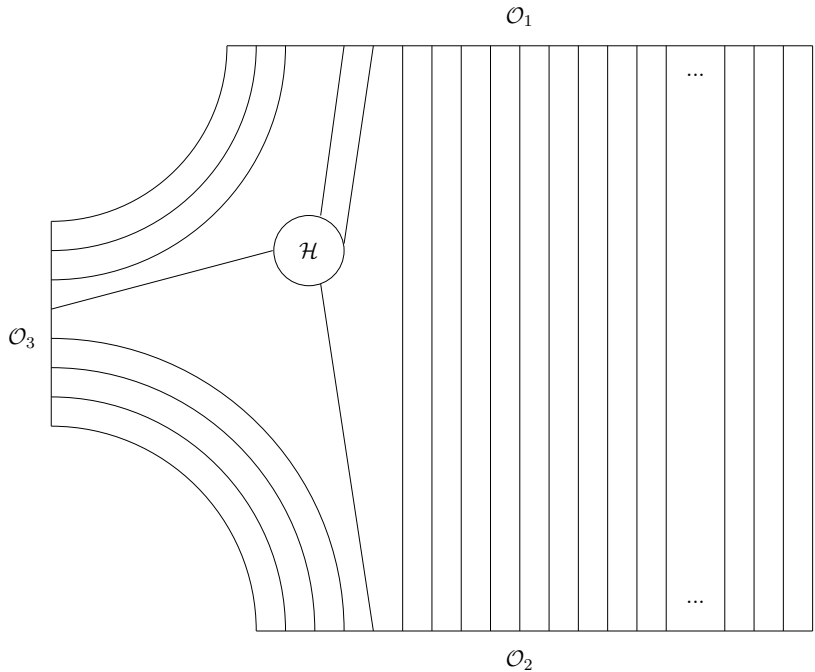
\begin{figure}
\begin{center}
\resizebox{11cm}{!}{
\begin{tikzpicture}
\draw (0.5,0) to (10,0);
\draw (0,10) to (10,10);
\draw (-3,3.5) to (-3,7);
\draw[](-3,7)arc(270:360:3cm);
\draw[](0.5,0)arc(0:90:3.5cm);
\draw[](1,0)arc(0:90:4cm);
\draw[](1.5,0)arc(0:90:4.5cm);
\draw[](2,0)arc(0:90:5cm);
\draw[](-3,6.5)arc(270:360:3.5cm);
\draw[](-3,6)arc(270:360:4cm);
\draw (1.4,6.5) circle (0.6cm);
\draw (-3,5.5) to (0.782,6.5);
\draw (8,0.5) node(x)  {...};
\draw (8,9.5) node(x)  {...};
\draw (1.4,6.5) node(x)  {$\mathcal{H}$};
\draw (2,10) to (1.6,7.1);
\draw (2.5,10) to (2,6.62);
\draw (2.5,0) to (1.6,5.93);
\draw (3,0) to (3,10);
\draw (3.5,0) to (3.5,10);
\draw (4,0) to (4,10);
\draw (4.5,0) to (4.5,10);
\draw (5,0) to (5,10);
\draw (5.5,0) to (5.5,10);
\draw (6,0) to (6,10);
\draw (6.5,0) to (6.5,10);
\draw (7,0) to (7,10);
\draw (7.5,0) to (7.5,10);
\draw (8.5,0) to (8.5,10);
\draw (9,0) to (9,10);
\draw (9.5,0) to (9.5,10);
\draw (10,0) to (10,10);
\draw (5,-0.5) node(x)  {$\mathcal{O}_2$};
\draw (5,10.5) node(x)  {$\mathcal{O}_1$};
\draw (-3.5,5) node(x)  {$\mathcal{O}_3$};
\end{tikzpicture}
}
\end{center}
\caption{\label{fig:loop} Example of a diagram contributing at one-loop with the insertion of the one-loop Hamiltonian with two legs in $\CO_3$ and the other two legs in $\CO_1$ and $\CO_2$ respectively. }
\end{figure}

At one loop there are two types of corrections that one should take into account. The first type is due to the two loop contribution to the effective sigma model description which amounts to corrections of order $\lambda$ to the external wave function. The second correction is due to the one-loop diagrams with two legs in one of the operators and the other two legs in two different operators, as shown in Fig~\ref{fig:loop}. These diagrams can be computed in the planar limit using the prescription of~\cite{Okuyama:2004bd, Roiban:2004va,Alday:2005nd}.

%

%

\subsubsection*{Two loop correction to the eigenstates}

The first type of correction has been neglected in earlier studies of 3-point functions of gauge theory operators in $\mathcal{N}=4$ SYM theory~\cite{Okuyama:2004bd, Roiban:2004va,Alday:2005nd}, as pointed out in~\cite{Beisert:2002bb, Kristjansen:2010kg, Grossardt:2010xq}.  While in general it is rather complicated to take into account this contribution, it actually becomes very easy for the particular set of operators that we are considering. This is due to the enormous simplification that one has by using a coherent state representation for the gauge theory operators. As already noticed in~\cite{Escobedo:2011xw}, this is also the reason that made the computation of the leading order contribution to $C_{123}$ possible. 

In brief, to take into account this type of contribution we should simply use in our expressions the wave function which is solution of the EOMs up to two-loops that one derives from \eqref{LLlagrg}, with a change of notation from the vector ${\vec n}$ to ${\bf u}$. In fact writing
\begin{equation}
{\bf u}={\bf u}^{(0)}+\lambda' {\bf u}^{(1)}+\CO(\lambda'^2)
\label{uexp}
\end{equation}
and substituting the full {\bf u} in \eqref{newescobedoC123b} one can compute these type of corrections order by order in $\lambda'$. We will implicitly compute these contributions by assuming that the function entering in the one-loop result for $C_{123}$ is the one in \eqref{uexp}.
This procedure can be extended to include also higher orders in $\lambda'$. Note here that we assume that Eq.~\eqref{vu_app} holds also at order $\lambda'$ since otherwise the difference between ${\bf u}$ and ${\bf v}$ would enter at order $\lambda'$ when inserting \eqref{uexp} in \eqref{newescobedoC123b}.


\subsubsection*{One-loop diagram}

The other type of correction contributing at one-loop comes form the insertion of the one-loop Hamiltonian with two legs in one of the operators and the other two legs in two different operators (see Fig~\ref{fig:loop}). 
We compute these corrections using the prescription given in~\cite{Okuyama:2004bd, Roiban:2004va,Alday:2005nd}. 
Since we have three operators, there are three types of diagrams.
Following~\cite{Okuyama:2004bd, Roiban:2004va,Alday:2005nd} we have that
\begin{equation}
C_{123}^{(1)}=\frac{1}{32 \pi^2} \frac{J^2 \CN_3}{N} B \sum_{k=1}^J  \left(  \frac{(\bar{u}^1  v_2 ) (\frac{k}{l} )}{(\bar{{\bf u}}  \cdot {\bf v} ) (\frac{k}{l} )} \right)^j\left(f^1_{23}(k)+f^2_{31}(k)+f^3_{12}(k)\right)
\label{okuyama}
\end{equation}
where $B$ is given in \eqref{b}, $f^1_{23}$ is the constant referring to the 3-point Feynman diagram with two contractions in ${\CO}_1$ and one contraction each with $\CO_2$ and $\CO_3$ and so on. 

For a given $k$, we compute 
\begin{equation}
f^{1}_{23} (k) = -  \frac{\bar{u}^{i_1} (\frac{k+j+1}{l} ) \bar{u}^{i_2} (\frac{k+j}{l} )  v_{j_1} (\frac{k+j+1}{l}) \delta^1_{j_2}}{ ({\bf \bar u}  \cdot {\bf v}) (\frac{k+j+1}{l} )\bar{u}^1 (\frac{k+j}{l} ) } \CH_{i_1i_2}^{j_1j_2}     
-  \frac{\bar{u}^{i_1} (\frac{k+1}{l} ) \bar{u}^{i_2} (\frac{k}{l} ) \delta^1_{j_1} v_{j_2} (\frac{k}{l}) }{\bar{u}^1 (\frac{k+1}{l} ) ({\bf \bar u} \cdot {\bf v}) (\frac{k}{l} ) } \CH_{i_1i_2}^{j_1j_2}
\end{equation}
\begin{equation}
f^{2}_{31} (k) = -  \frac{\delta_{2}^{i_1} \bar{u}^{i_2} (\frac{k+j+1}{l} ) v_{j_1} (\frac{k+j}{l})  v_{j_2} (\frac{k+j+1}{l}) }{v_2 (\frac{k+j}{l} )({\bf \bar u}  \cdot {\bf v}) (\frac{k+j+1}{l} ) } \CH_{i_1i_2}^{j_1j_2}  -  \frac{ \bar{u}^{i_1} (\frac{k}{l} ) \delta_{2}^{i_2} v_{j_1} (\frac{k}{l})  v_{j_2} (\frac{k+1}{l}) }{({\bf \bar u} \cdot {\bf v}) (\frac{k}{l} ) v_2 (\frac{k+1}{l} )  } \CH_{i_1i_2}^{j_1j_2} 
\end{equation}
with
\begin{equation}
\label{interaction}
\CH_{i_1i_2}^{j_1j_2 }=2(I-P)_{i_1, i_2}^{j_1, j_2},\spa I_{i_1 i_2}^{j_1 j_2} = \delta_{i_1}^{j_1} \delta_{i_2}^{j_2} \spa P_{i_1 i_2}^{j_1 j_2} = \delta_{i_1}^{j_2} \delta_{i_2}^{j_1}
\end{equation}
From our choice of the operator $\CO_3$, one can see that $f^{3}_{12}=0$.
This is because, among all the states in $\CO_3$, only two contribute to $f^3_{12}$, namely $\tr\left({\bar X}^j Z^j\right)$ and $\tr\left({\bar X}^{j-1}Z{\bar X} Z^{j-1}\right)$ with a relative minus sign.
 
Using that ${\bf \bar u}(\sigma) $ and ${\bf v}(\sigma)$ vary slowly, along with Eq.~\eqref{gaugevir} and Eq.~\eqref{vu_app},  we compute
\begin{eqnarray}
-\frac{1}{2} f^{1}_{23} (k)   &=&   2       -      \frac{\bar{u}^{1} (\frac{k+j+1}{l} ) }{\bar{u}^1 (\frac{k+j}{l} )}        \frac{{\bf \bar u} (\frac{k+j}{l} ) \cdot  {\bf v} (\frac{k+j+1}{l}) }{  ({\bf \bar u}  \cdot {\bf v}) (\frac{k+j+1}{l} )}       -  \frac{\bar{u}^1 (\frac{k}{l} )}{\bar{u}^1 (\frac{k+1}{l} )}     \frac{{\bf \bar u} (\frac{k+1}{l} ) \cdot {\bf v} (\frac{k}{l}) }{   ({\bf \bar u} \cdot {\bf v}) (\frac{k}{l} )} \nn \\ &=& 2 - \left. \left( 1 + \frac{1}{l} \frac{\bar{u}^1{}'}{\bar{u}^1} + \frac{1}{2l^2} \frac{\bar{u}^1{}''}{\bar{u}^1} \right) \right|_{\sigma=\frac{k+j}{l}} \left. \left( 1 - \frac{1}{l} \frac{{\bf \bar u}' \cdot {\bf v}}{{\bf \bar u}\cdot {\bf v}} + \frac{1}{2l^2} \frac{{\bf \bar u}'' \cdot {\bf v}}{{\bf \bar u}\cdot {\bf v}} \right) \right|_{\sigma=\frac{k+j+1}{l}} \nn \\ && - \left. \left( 1 - \frac{1}{l} \frac{\bar{u}^1{}'}{\bar{u}^1} + \frac{1}{2l^2} \frac{\bar{u}^1{}''}{\bar{u}^1} \right) \right|_{\sigma=\frac{k+1}{l}} \left. \left( 1 + \frac{1}{l} \frac{{\bf \bar u}' \cdot {\bf v}}{{\bf \bar u}\cdot {\bf v}} + \frac{1}{2l^2} \frac{{\bf \bar u}'' \cdot {\bf v}}{{\bf \bar u}\cdot {\bf v}} \right) \right|_{\sigma=\frac{k}{l}} \nn \\ &=& \left\{ 2 - \left( 1 + \frac{1}{l} \frac{\bar{u}^1{}'}{\bar{u}^1}+ \frac{j}{l^2} \Big(\frac{\bar{u}^1{}'}{\bar{u}^1} \Big)' + \frac{1}{2l^2}  \frac{\bar{u}^1{}''}{\bar{u}^1} \right) \left( 1 - \frac{j}{l^2} {\bf \bar u}' \cdot \delta {\bf u} + \frac{1}{2l^2} {\bf \bar u}'' \cdot {\bf u} \right) \right. \nn \\ && \left. \left. - \left( 1 - \frac{1}{l} \frac{\bar{u}^1{}'}{\bar{u}^1}- \frac{1}{l^2} \Big(\frac{\bar{u}^1{}'}{\bar{u}^1} \Big)' + \frac{1}{2l^2}  \frac{\bar{u}^1{}''}{\bar{u}^1} \right) \left( 1 + \frac{j}{l^2} {\bf \bar u}' \cdot \delta {\bf u} + \frac{1}{2l^2} {\bf \bar u}'' \cdot {\bf u} \right) \right\} \right|_{\sigma=\frac{k}{l}} \nn \\ &=& \left. \frac{1}{l^2} \left\{   {\bf \bar u}' \cdot {\bf u}'  - (j-1) \Big(\frac{\bar{u}^1{}'}{\bar{u}^1} \Big)' -  \frac{\bar{u}^1{}''}{\bar{u}^1} \right\} \right|_{\sigma=\frac{k}{l}}
\end{eqnarray}
where we included terms up to order $1/J^2$. Similarly, we find
\begin{eqnarray}
-\frac{1}{2} f^{3}_{12} (k)&=&2        -  \frac{v_{2} (\frac{k+j+1}{l} )}{v_2 (\frac{k+j}{l} )}         \frac{{\bf \bar u} (\frac{k+j+1}{l} ) \cdot  {\bf v} (\frac{k+j}{l})  }{ ({\bf \bar u}  \cdot {\bf v}) (\frac{k+j+1}{l} ) }       -       \frac{v_2 (\frac{k}{l} )}{v_2 (\frac{k+1}{l} )}        \frac{{\bf \bar u} (\frac{k}{l} ) \cdot {\bf v} (\frac{k+1}{l}) }{ ({\bf \bar u} \cdot {\bf v}) (\frac{k}{l} ) } 
 \nn \\ &=& \left. \frac{1}{l^2} \left\{   {\bf \bar u}' \cdot {\bf u}'  - (j-1) \Big(\frac{u_2{}'}{u_2} \Big)' -  \frac{u_2{}''}{u_2} \right\} \right|_{\sigma=\frac{k}{l}}
\end{eqnarray}
Inserting these results in \eqref{okuyama} we obtain
\begin{equation}
\label{C123oneloop}
C_{123}^{(1)} = -\frac{1}{2\,N}\frac{  j! J }{ \sqrt{ (2j-1)! } } \int_0^{2\pi} \frac{d\sigma}{2\pi} ( \bar{u}^1 u_2 )^{j}  \left\{  {\bf \bar u}' \cdot {\bf u}'-\frac{j-1}{2} \Big( \frac{\bar{u}^1{}'}{\bar{u}^1} + \frac{u_2{}'}{u_2} \Big)' - \frac{1}{2} \Big( \frac{\bar{u}^1{}''}{\bar{u}^1} +  \frac{u_2{}''}{u_2}  \Big) 
\right\}
\end{equation}
Combining this with the result for the leading order \eqref{newescobedoC123b} with the wave function $\bf u$ solution of the EOMs up to two loops, we thus arrive at the final expression for the 3 point function \eqref{3pointfct}
\begin{equation}
\label{gaugeC123}
C_{123} =  \frac{  j! J }{N \sqrt{ (2j-1)! } } \int_0^{2\pi} \frac{d\sigma}{2\pi} ( \bar{u}^1 u_2 )^{j} \left[ 1-\frac{\lambda'}{2} \left\{ {\bf \bar u}' \cdot {\bf u}' + \frac{j^2-1}{2} \Big( \frac{(\bar{u}^1 u_2)'}{\bar{u}^1 u_2}  \Big)^2  +  \frac{\bar{u}^1{}' u_2'}{\bar{u}_1 u_2} 
\right\}
\right]  + \CO(\lambda^2)
\end{equation}
where we used partial integration to remove double derivatives.
In Section \ref{string} we compute the holographic dual of this quantity.

\subsection{Correction from spin-flipped coherent state}
\label{sec:spinflip}

In the above, we computed the one-loop correction to the 3-point function for two heavy operators $\CO_1 (x_1)$ and $\CO_2 (x_2)$ and one light chiral primary operator $\CO_3 (x_3)$ using the coherent state description \eqref{CO1} and \eqref{CO2} for the two heavy operators. However, as found in \cite{Kruczenski:2004kw}, while at order $\lambda$ gauge theory operators can be described in the long-wave length approximation using a coherent state, at order $\lambda^2$ one has to use a linear combination of a coherent state and a spin-flipped coherent state. This arises when integrating out the short scale degrees of freedom in the spin chain description. We consider below the effect of using the full linear combination, instead of only the coherent state part that we using in \eqref{CO1}-\eqref{CO2}.

\subsubsection*{Correction to coherent state description from spin-flipped coherent state}

Consider first the coherent state part. Note that we work in the $SU(2)$ sector in the following. We represent this by the state
\begin{equation}
\label{psi0}
| \psi_0 \rangle = | \vec{n}_1 \rangle \otimes | \vec{n}_2 \rangle \otimes \cdots  \otimes | \vec{n}_J \rangle 
\end{equation}
where for each site we write $| \vec{n}_k \rangle = R_k \, | \!  \!  \uparrow \rangle$ with $R_k$ being a rotation matrix for the $k$'th site. The continuum description uses instead the function $\vec{n} ( \frac{2\pi k}{J} ) = \vec{n}_k$.
The state \eqref{psi0} corresponds to the description of the $\CO_1 (x_1)$ and $\CO_2 (x_2)$ operators using Eqs.~\eqref{CO1}-\eqref{CO2}. Note also that we require $\vec{n} (\sigma)$ to solve the EOMs of the two-loop effective Lagrangian \eqref{LLlagrg}.

However, as found in \cite{Kruczenski:2004kw}, the full gauge theory state at order $\lambda^2$ (and for large $J$) is given by
\begin{equation}
\label{corrstate}
|\psi \rangle = \Big(  1 - \frac{1}{2} \sum_{k,k'} | c_{k,k'} |^2 \Big) | \psi_0 \rangle + |\psi_1 \rangle \spa |\psi_1 \rangle =  \sum_{k,k'} c_{k,k'} |k,k' \rangle
\end{equation}
where $|k,k'\rangle$ is built from the coherent states with two spin flips
\begin{equation}
| \!  \downarrow_a \downarrow_b \rangle = R_1 |\! \! \uparrow \rangle \otimes \cdots \otimes R_{a-1} | \! \! \uparrow\rangle \otimes R_a |\! \! \downarrow \rangle \otimes \cdots \otimes R_{b-1} |\! \!  \uparrow\rangle \otimes R_b |\! \! \downarrow \rangle \otimes \cdots \otimes R_J |\! \! \uparrow \rangle
\end{equation}
as follows
\begin{equation}
| k , k' \rangle = \frac{\sqrt{2}}{J} e^{-i(k+k')p} \sum_{b > a=1}^J e^{ika+ik'b} |\! \downarrow_a \downarrow_b \rangle 
\end{equation}
where $p$ is a number giving an optional extra phase factor. Using the results and notation of \cite{Kruczenski:2004kw} we can write
\begin{equation}
\label{prepsi1}
|\psi_1 \rangle = \frac{J}{2\sqrt{2}}  \sum_a \sum_{k,k'} \sum_{q=1}^2 \lambda_q  \frac{ e^{i(k+k')(a-p)}e^{ik'q}}{\epsilon(k) + \epsilon(k') }  A_{--}^{a,a+q}  |\! \downarrow_a \downarrow_{a+q} \rangle  
\end{equation}
with
\begin{equation}
\lambda_1 = \frac{1}{4\pi^2} - \frac{\lambda}{16\pi^4} \spa \lambda_2 = \frac{\lambda}{64\pi^4} \spa \epsilon(k) = J^2 [ \lambda_1 (1-\cos k) + \lambda_2 (1-\cos 2k ) ]
\end{equation}
where $\epsilon (k)$ is the energy for one spin flip, and for large $J$ we have
\begin{equation}
A_{--}^{a,a+q} \simeq \frac{1}{2} \Big( \frac{2\pi q}{J} \Big)^2 B ( \frac{2\pi a}{J} )
\spa B(\sigma) = - (\partial_\sigma \theta)^2 + \sin^2 \theta (\partial_\sigma \varphi)^2 - 2i \sin \theta \partial_\sigma \theta \partial_\sigma \varphi
\end{equation}
We now extract the part of this proportional to $\lambda$, discarding the terms which either give finite-size corrections at order $\lambda^0$ or terms of order $\lambda^2$. Then, for large $J$, we can write
\begin{equation}
\label{psi1}
|\psi_1 \rangle = \frac{\lambda'}{4\sqrt{2}} \sum_a B ( \frac{2\pi a}{J} ) \sum_{q=1}^2 F_q(a-p) |\! \downarrow_a \downarrow_{a+q} \rangle  
\end{equation}
\begin{equation}
\label{Fdef}
F_1(a) = -  \frac{1}{4J} \sum_{k,k'} \frac{e^{i(k+k')a}e^{ik'}(2-\cos 2k - \cos 2k')}{(2-\cos k - \cos k')^2}   \spa
F_2(a) = \frac{1}{J} \sum_{k,k'} \frac{e^{i(k+k')a} e^{2ik'}}{2-\cos k - \cos k'}  
\end{equation}

\subsubsection*{Spin-flip correction and the 3-point function}

We now turn to the impact on the 3-point function computed in this paper. We can schematically write the two heavy operators as $\CO_i(x_i) = \CO_i^{(0)} (x_i) + \CO_i^{(1)} (x_i)$, $i=1,2$, where $\CO_i^{(0)} (x_i)$ are now the operators given in Eqs.~\eqref{CO1}-\eqref{CO2} using a coherent state description, and $\CO_i^{(1)} (x_i)$ are the spin-flip corrections which can be inferred from \eqref{prepsi1}. Note that to one-loop order we can approximate $|\psi \rangle = |\psi_0\rangle + |\psi_1\rangle$. We have
\begin{eqnarray}
\label{sf3p}
&& \langle {\mathcal{O}}_1(x_1){\mathcal{O}}_2(x_2){\mathcal{O}}_3(x_3)\rangle = \langle {\mathcal{O}}_1^{(0)}(x_1){\mathcal{O}}_2^{(0)}(x_2){\mathcal{O}}_3(x_3)\rangle + \langle {\mathcal{O}}_1^{(1)}(x_1){\mathcal{O}}_2^{(0)}(x_2){\mathcal{O}}_3(x_3)\rangle \nn \\ && + \langle {\mathcal{O}}_1^{(0)}(x_1){\mathcal{O}}_2^{(1)}(x_2){\mathcal{O}}_3(x_3)\rangle + \langle {\mathcal{O}}_1^{(1)}(x_1){\mathcal{O}}_2^{(1)}(x_2){\mathcal{O}}_3(x_3)\rangle 
\end{eqnarray}
We first remark that to one-loop order, we can only get a contribution from the spin-flip correction for the tree-level diagram since $|\psi_1\rangle$ in Eq.~\eqref{psi1} is proportional to $\lambda$. Hence this also holds for $\CO_{1,2}^{(1)} (x_i)$. From this it is also clear that the last term on the RHS of Eq.~\eqref{sf3p} is of order $\lambda^2$. Thus, at one-loop, the possible contributions from the spin-flipped coherent state corrections can come from computing the tree-level Wick contractions for the second and third terms on the RHS of Eq.~\eqref{sf3p} at tree-level. 

Consider the Wick contractions of the term $\langle {\mathcal{O}}_1^{(0)}(x_1){\mathcal{O}}_2^{(1)}(x_2){\mathcal{O}}_3(x_3)\rangle$. Thus, while ${\mathcal{O}}_1^{(0)}(x_1)$ is inferred from $|\psi_0\rangle$ the operator ${\mathcal{O}}_2^{(1)}(x_2)$ is inferred from $|\psi_1\rangle$. Consider now $|\psi_1\rangle$ of Eq.~\eqref{psi1}. Considering the tree-level Wick contractions we see that if the index $a$ in \eqref{psi1} points to a site that contracts with $\CO_1^{(0)}(x_1)$, the contribution is zero since $\langle \uparrow \! | \! \downarrow \rangle = 0$. Hence the non-zero contribution comes from values of $a$ that point to sites that contracts with $\CO_3(x_3)$ (and also such that either $a+1$ or $a+2$ contracts with $\CO_3(x_3)$). Each of these sites contracts with an $\bar{X}$ in $\CO_3(x_3)$. Due to the two spin flips, the contraction with the operator corresponding to the state $|\! \downarrow_a \downarrow_{a+q} \rangle$ picks up a factor $u_2^{j-2} u_1^2$ which combined with the Wick contractions between $\CO_1(x_1)$ and $\CO_3(x_3)$ gives a factor $(\bar{u}^1 u_2)^{j-2} (\bar{u}^1 u_1)^2$ factor. Combined with the other parts of \eqref{psi1} we pick up the contribution
\begin{equation}
\label{flipped}
\frac{\lambda'}{4\sqrt{2}} B(\sigma) \left[ \sum_{a=1}^{j-1} F_1 (a-p) + \sum_{a=1}^{j-2} F_2 (a-p) \right]
\end{equation}
One can find numerically that $F_1(a) + F_2(a)$ is of order $1/J$ for $a \neq 0$. However, taken separately $F_{1,2}(a)$ are of order $J$. Moreover, $F_{1,2} (a)$ peaks around $a=0$. Indeed for large $J$ one finds that $F_1(a) + F_2(a) \propto \delta_a$ (note that this result is consistent with using the approximation $\epsilon(k) + \epsilon(k') \simeq 2\epsilon(k')$ in \cite{Kruczenski:2004kw}). Hence, the contribution \eqref{flipped} is highly sensitive to the value of $p$. Since we seem to end up with a divergent result, we leave a further and more careful analysis of the correction from the spin-flipped coherent state to future investigation. 

However, we end this section with the following remarks. First, from conservation of R-charge one could argue that the spin flip correction should end up being zero. Indeed, the expectation values of the R-charges changes when flipping the spins in the coherent state. This suggests that by R-charge conservation the second and third terms on the RHS \eqref{sf3p} should be zero. However, this is not a precise argument since the coherent states are not eigenstates of the R-charges. Nevertheless one could speculate that the fact that the R-charges in $\langle {\mathcal{O}}_1^{(1)}(x_1){\mathcal{O}}_2^{(0)}(x_2){\mathcal{O}}_3(x_3)\rangle$ and $\langle {\mathcal{O}}_1^{(0)}(x_1){\mathcal{O}}_2^{(1)}(x_2){\mathcal{O}}_3(x_3)\rangle$ are not conserved on the level of expectation values should mean that their contributions are highly suppressed in the large $J$ limit.

On a further note, a possible contribution from the spin-flip correction would seem to be proportional to the function $(\bar{u}^1 u_2)^{j-2} ( \bar{u}^1 u_1 B(\sigma) + \bar{u}^2 u_2 \bar{B}(\sigma) )$. This function is not proportional to any of the three terms at order $\lambda$ in \eqref{gaugeC123}. Thus, if this contribution is non-zero it would seem that it introduces a new type of term in the 3-point function coefficient \eqref{gaugeC123}.


\section{String theory side}
\label{string}

In this section we describe the computation of the one- and two-loop correction to the holographic 3-point function coefficient for the case of the two semiclassical operators and the small $1/2$ BPS operator considered in the previous section. This is done following the work initiated in Ref.~\cite{Zarembo:2010rr, Escobedo:2011xw}.  The two large operators are described by semiclassical strings while the small BPS operator corresponds to a quantum string. 

Our starting point is the sigma-model for type IIB string theory on $\ads_5\times S^5$ in the regime in which it is described by the Landau-Lifshitz sigma-model~\cite{Kruczenski:2003gt}.

We use in the following that
\begin{equation}
\label{R4} 
R^4 =  \lambda (\alpha')^2
\end{equation}
This relates the string parameters $R$ and $\alpha'$ to the 't Hooft coupling $\lambda$ of $\CN=4$ SYM.
Using this we can formulate the string theory result in terms of gauge theory variables.

In the following we show how to compute the sigma-model Lagrangian up to the order $\lambda^3$. This is because we want to compute the 3-point correlation function coefficient up to $\CO(\lambda^2)$. As anticipated in the Introduction, to do this one should use the wave function which is solution of the EOMs up to $\CO(\lambda^3)$, since at each order in the expansion parameter, the wave function receives corrections coming from the next order contribution to the effective sigma model description. The explicit computation of the sigma model Lagrangian is performed explicitly up to and including $\CO(\lambda^2)$ corrections. For the contribution at $\CO(\lambda^3)$ we explain in words the procedure and we report the result~\cite{Kruczenski:2004kw, Minahan:2005qj}.~\footnote{An analogous computation up to the same order in $\lambda$ has been performed in detail in~\cite{Grignani:2008is, Astolfi:2008ji} for the case of the $\ads_4/CFT_3$ correspondence.} At the end of this section we compare the result that we get just at one-loop on the string side with the corresponding quantity computed on the gauge theory side.

In the rest of this section we set $\alpha' = 1$ for simplicity. The metric for type IIB string theory on $\ads_5 \times S^5$ can be written as
\begin{equation}
\label{metads} ds^2 = R^2 \left[ - \cosh^2 \rho \, dt^2 + d\rho^2 +
\sinh^2 \rho \, (d\Omega_3')^2 + d\zeta^2 + \sin^2 \zeta \,
d\alpha^2 + \cos^2 \zeta \, (d\Omega_3)^2 \right]
\end{equation}
and we also have the five-form Ramond-Ramond field strength
\begin{equation}
\label{fsads} F_{(5)} = 2 R^4 \left[ \cosh \rho \, \sinh^3 \rho \,
dt \, d\rho \, d\Omega_3' + \sin \zeta \, \cos^3\zeta \, d\zeta \,
d\alpha \, d\Omega_3 \right]
\end{equation}
The three-sphere $\Omega_3$ is parametrized as
\begin{equation}
\label{threesphere} (d\Omega_3)^2 = d\psi^2 + \cos^2 \psi d\phi_1^2
+ \sin^2 \psi d\phi_2^2 = d\psi^2 + d\phi_-^2 + d\phi_+^2 + 2 \cos
(2\psi) d\phi_- d\phi_+
\end{equation}
where $2 \phi_\pm = \phi_1 \pm \phi_2$. The energy $E$ of a string state and the
$SO(6)$ Cartan generators $J_i$, $i=1,2,3$, are given by
\begin{equation}
E = i \partial_t \spa J \equiv J_1 + J_2 = - i \partial_{\phi_+}
\spa J_3 = - i
\partial_\alpha
\end{equation}
The holographic dual of the 3-point correlation function studied in the previous section corresponds to considering a string which is point like in $AdS$ and is moving non trivially on a 3-sphere contained in $S^5$, namely we consider the classical sigma-model on $\R\times S^3$, therefore we are only interested in the charges $E$, $J_1$ and $J_2$ and we can restrict to the region
$\rho=\zeta =0$. This gives the metric $ds^2 = R^2 [ -dt^2 + (d\Omega_3)^2]$.
Introducing the new angles
\begin{equation}
\label{thetaphi}
\theta \equiv 2 \psi - \frac{\pi}{2} \spa \varphi \equiv 2 \phi_-
\end{equation}
the metric becomes
\begin{equation}
\label{theRxS3}
ds^2 = R^2 \left[ - dt^2 + \frac{1}{4} (d\Omega_2)^2 + \left( d
\phi_+ + \frac{1}{2} \sin \theta d\varphi \right)^2 \right]
\spa
(d\Omega_2)^2 =  d\theta^2 + \cos^2 \theta d\varphi^2
\end{equation}

It is convenient to
 introduce the coordinates
\begin{equation}
\label{lccoor}
x^+ = \lambda' t \spa x^- = \phi_+ - t
\end{equation}
where $\lambda'=\lambda/J^2$ and we are considering the limit $\lambda'\rightarrow 0$ as in the gauge theory case. In these coordinates we have the following identification of the charges
\begin{equation}
i\partial_+={H}=\frac{E-J}{ \lambda'} \spa -i\partial_-=J
\label{relcharges}
\end{equation}
The metric then takes the form
\begin{equation}
ds^2 = R^2 \Big[ \frac{1}{4} (d\Omega_2)^2 + \Big( 2 \frac{dx^+}{\lambda'}+ dx^- + \omega \Big) ( dx^- + \omega) \Big]\spa \omega=\frac{1}{2} \sin \theta d\varphi 
\end{equation}
The bosonic sigma-model Lagrangian and the Virasoro constraints are respectively
\begin{equation}
\CL = - \frac{1}{2} h^{\alpha\beta} G_{\mu\nu} \partial_\alpha x^\mu
\partial_\beta x^\nu
\end{equation}
\begin{equation}
G_{\mu\nu} ( \partial_\alpha x^\mu \partial_\beta x^\nu -
\frac{1}{2} h_{\alpha\beta} h^{\gamma\delta} \partial_\gamma x^\mu
\partial_\delta x^\nu ) = 0
\end{equation}
where $h^{\alpha\beta}= \sqrt{-\det \gamma} \gamma^{\alpha\beta}$ with $\gamma_{\alpha\beta}$ being the world-sheet metric. 
We define for convenience
\begin{equation}
A \equiv - h^{00} \spa B \equiv h^{01}\spa
S_{\alpha\beta} \equiv G_{\mu\nu} \partial_\alpha x^\mu \partial_\beta x^\nu
\end{equation}
where we used that $h^{\alpha\beta}$ has only two independent components since $\det h = -1$, thus $h^{11}=(1-B^2)/A$.
The Lagrangian and Virasoro constraints can now be written as
\begin{equation}
\CL = \frac{A}{2} S_{00} - B S_{01} - \frac{1-B^2}{2A} S_{11}
\end{equation}
\begin{equation}
\label{vircon} \begin{array}{c} \ds (1+B^2) S_{00} +
\frac{2B(1-B^2)}{A} S_{01} + \frac{(1-B^2)^2}{A^2} S_{11} = 0
\\[4mm] \ds
AB S_{00} + 2(1-B^2) S_{01} - \frac{B(1-B^2)}{A} S_{11} =0
\end{array}
\end{equation}
We make the following gauge choice 
\begin{equation}
\label{choice} 
x^+ = \kappa \tau
\end{equation}
\begin{equation}
\label{gaugecon} 
2\pi  p_- = \frac{\partial \CL}{\partial
\partial_\tau x^-} = \mbox{const.} \spa \frac{\partial \CL}{\partial
\partial_\sigma x^-} = 0
\end{equation}
where $\kappa$ is a constant.

From \eqref{relcharges} we see that $\tau$ does not give the right energy scale on the world-sheet.
Therefore we introduce $\tilde\tau=\kappa\tau$ and use the notation
\begin{equation}
\dot{x}^\mu = \frac{\partial x^\mu}{\partial {\tilde\tau}} \spa (x^\mu)' = \frac{\partial x^\mu}{\partial \sigma}
\end{equation}
We moreover make the following expansions of the quantities $A$ and $B$
\begin{equation}
A = 1 +{\kappa^2} A_1 +{\kappa^4} A_2 +\cdots \spa B ={\kappa^3} B_1 +
{\kappa^5} B_2 + \cdots
\end{equation}
This is consistent with the fact that to leading order 
we have that $A=1$ and $B=0$.
We can then determine the constant $\kappa$ from \eqref{gaugecon} and to leading order in $\lambda'$ we find 
\begin{equation}
J = \int_0^{2\pi} d\sigma p_- =\frac{ R^2 \kappa}{\lambda'}
\end{equation}
Therefore, using~\eqref{R4}, we have that $\kappa = \sqrt{\lambda'}$.
We see thus that $\kappa \rightarrow 0$.~\footnote{Note that the constant $\kappa$ in this paper corresponds to $1/\kappa$ of Ref.~\cite{Escobedo:2011xw}.} 

We can now solve the gauge conditions as
\begin{equation}
\label{xminussol}
\dot{x}^- = - \frac{1}{2} \sin \theta \dot{\varphi} - A_1 + {\kappa^2} (A_1^2 - A_2) + \CO( \kappa^4 ) \spa
{x^-}' = - \frac{1}{2} \sin \theta \varphi' - {\kappa^2}B_1+ \CO( \kappa^4 ) 
\end{equation}
Inserting this in the Virasoro constraints we can now find the solution for $A_1$, $A_2$ and $B_1$
\begin{equation}
\label{theAB1}
A_1 = \frac{1}{8} ( {\theta'}^2 + \cos^2 \theta {\varphi'}^2 )
\spa
B_1 = \frac{1}{4} ( \dot{\theta} \theta' + \cos^2 \theta \dot{\varphi} \varphi' )
\end{equation}
\begin{equation}
\label{theA2}
A_2 = \frac{1}{8} ( \dot{\theta}^2 + \cos^2 \theta \dot{\varphi}^2 ) - \frac{1}{128} ( {\theta'}^2 + \cos^2 \theta {\varphi'}^2 )^2
\end{equation}
To write the gauge fixed Lagrangian
\begin{equation}
\CL_g = \CL - 2\pi  \kappa p_-  \dot{x}^-
\end{equation}
we plug in $\dot{x}^-$, ${x^-}'$, $A$ and $B$ from
\eqref{xminussol} and \eqref{theAB1}-\eqref{theA2}. 
This gives an expansion in powers of $\lambda'$
\begin{equation}
\CL_g = \CL_0 +{\lambda'}\CL_1  + \cdots
\end{equation}
with 
\begin{equation}
\label{e1}
\frac{1}{R^2} \CL_0 =  \frac{1}{2} \sin \theta \dot{\varphi} - \frac{1}{8} ( {\theta'}^2 + \cos^2 \theta {\varphi'}^2 )
\end{equation}
\begin{equation}
\label{e2}
\frac{1}{R^2} \CL_1 = \frac{1}{8} ( \dot{\theta}^2 + \cos^2 \theta \dot{\varphi}^2 ) + \frac{1}{128} ( {\theta'}^2 + \cos^2 \theta {\varphi'}^2 )^2
\end{equation}
From $\CL_0$, one gets the energy at the order $\lambda'$ as can be seen from~\eqref{relcharges}. From $\CL_1$ therefore one obtains the energy at order $\lambda'^2$ and so on.

To compute the two-loop correction to the 3-point correlation function coefficient we have to solve the Landau-Lifshiz sigma model to order $\lambda'^3$. This in fact would give a contribution of order $\lambda'^2$ to the wave function that appears in the final result. We  therefore include this contribution in the computation.

Here we only showed explicitly how to solve the sigma-model up to the order $\lambda'^2$, corresponding to $\CL_1$, but the computation can be easily extended to the next order.
However, it is convenient to introduce a more suitable notation in terms of the following parameterization
\begin{equation}
\vec{n} = ( \cos \theta \cos \varphi , \cos \theta \sin
\varphi , \sin \theta )
\end{equation}
where $\vec{n}$ is a unit vector pointing to a point in the two-sphere. Using this notation we can rewrite the expressions \eqref{e1} and \eqref{e2} as
\begin{equation}
\frac{1}{R^2} \CL_0 =  \frac{1}{2} \sin \theta \dot{\varphi} - \frac{1}{8} (\vec{n}')^2
\spa
\frac{1}{R^2} \CL_1 = \frac{1}{8} \dot{\vec{n}}^2 + \frac{1}{128} (\vec{n}')^4
\end{equation}
One should now make a field redefinition that removes the time derivatives in the Lagrangian. It has been shown \cite{Kruczenski:2004kw} that, at the order we are working, this field redefinition corresponds to evaluating $\CL_1$ on-shell, $i.e.$ to substitute in the solution of the EOMs from $\CL_0$ to get rid of the time derivatives. From $\CL_0$ we find the EOMs
\begin{equation}
2 \vec{n} \times \dot{\vec{n}} = - \vec{n}'' - \vec{n} ( \vec{n}' )^2
\end{equation}
We compute from this
\begin{equation}
4 \dot{\vec{n}}^2 = (\vec{n}'')^2 - (\vec{n}')^4
\end{equation}
Thus the on-shell $\CL_1$ is 
\begin{equation}
(\CL_1)_{\rm on} = \frac{1}{32} (\vec{n}'')^2 - \frac{3}{128} (\vec{n}')^4
\end{equation}
and the field redefined gauge fixed Lagrangian is
\begin{equation}
\CL_g = \CL_0 + \lambda'( \CL_1 )_{\rm on}  + \cdots
\end{equation}
giving
\begin{equation}
\frac{1}{R^2} \CL_g = \frac{1}{2} \sin \theta \dot{\varphi} - \frac{1}{8} (\vec{n}')^2 +  \frac{\lambda'}{32} \Big[ (\vec{n}'')^2 - \frac{3}{4} (\vec{n}')^4 \Big] + \CO (\lambda'^2 )
\end{equation}

We can now proceed in the same way and include the next order in the computation. Also in this case we should 
perform a field redefinition to remove the time derivative from the $\lambda'^2$ correction to the 
Lagrangian~\cite{Kruczenski:2004kw, Minahan:2005qj}.
The final result is
\begin{eqnarray}
\label{LLlagr}
\frac{1}{R^2} \CL_g &=& \frac{1}{2} \sin \theta \dot{\varphi} - \frac{1}{8} (\vec{n}')^2 +  \frac{\lambda'}{32} \Big[ (\vec{n}'')^2 - \frac{3}{4} (\vec{n}')^4 \Big] \cr
&-& \frac{\lambda'^2}{64} \Big[ (\vec{n}''')^2 - \frac{7}{4} (\vec{n}')^2(\vec{n}'')^2 - \frac{25}{2} (\vec{n}' \vec{n}'')^2 +\frac{13}{16} (\vec{n}')^6
\Big]+ \CO (\lambda'^3 )
\end{eqnarray}

\subsubsection*{3-point function at order ${\lambda'}$}

We now are ready to use this result to compute the corrections to the holographic 3-point correlation function coefficient $C_{123}$ for two semi-classical operators and a light chiral primary operator up to two-loops. The prescription for computing this coefficient was put forward in~\cite{Zarembo:2010rr} and in our notation becomes
\begin{equation}
\label{C123string}
C_{123} = c_j \frac{\sqrt{\lambda}}{N}\int_{-\infty}^{+\infty} d{\tau_e} \int_{0}^{2 \pi} \frac{d\sigma}{2\pi}\frac{ (\bar{U}^1 U_2 )^{j}  }{\cosh^{2j} (\frac{\tau_e}{\kappa})} \left[ \frac{2}{\kappa^2\cosh^2 (\frac{\tau_e}{\kappa})}-\frac{1}{\kappa^2}-\partial_a {\bf \bar U}\cdot\partial^a {\bf U}\right]
\end{equation}
where $\tau_e$ is the Euclidean time and we already used the gauge choice~\eqref{choice}.  $c_j$ is a constant depending only on the parameter $j$ which is associate to the supergravity mode dual to the chiral primary operator. In our case it is given by
\begin{equation}
\label{cj}
c_j=\frac{(2j+1)!}{2^{2j+2}j!\sqrt{(2j-1)!}}
\end{equation}
Here ${\bf U}(\tau,\sigma)$ is a complex vector that parametrizes the embedding of the type IIB string on $S^5$.  We have 
\begin{equation}
\label{Ucoord}
U_1=\sin\psi e^{i\phi_1} \spa U_2=\cos\psi e^{i\phi_2} \spa U_3=0
\end{equation}
and we work in the Frolov-Tseytlin limit~\cite{Frolov:2003xy, Kruczenski:2003gt} which in our notation is 
\begin{equation}
\label{TMlimit}
 \kappa \rightarrow 0 \spa\frac{1}{\kappa}\partial_{\tau} {\bf U}~~{\rm fixed}\spa \partial_\sigma {\bf U}~~{\rm fixed}
 \end{equation}
which provides an expansion in $ \lambda'$ which parallels the loop expansion on the gauge theory side. 
We can then compute the term $\partial_a {\bf \bar U}\cdot\partial^a {\bf U}$ appearing in \eqref{C123string}. In the limit \eqref{TMlimit} it becomes
\begin{eqnarray}
\partial_a {\bf \bar U}\cdot\partial^a {\bf U}&=&-\frac{1}{\kappa^2}+\frac{1}{2}(\vec{n}')^2-\frac{\kappa^2}{16}(\vec{n}')^4+\mathcal{O}(\kappa^4)
\end{eqnarray}
Therefore up to terms second order in $\kappa^2$ (or equivalently $\lambda'$)  we get 
\begin{multline}
\label{3ptdiff}
C_{123}=  c_j \frac{\sqrt{\lambda}}{N}\int_{-\infty}^{+\infty} d{\tau_e} \int_{0}^{2 \pi} \frac{d\sigma}{2\pi}\frac{ (e^{-i\varphi}\cos\theta)^{j}  }{2^j\cosh^{2j} (\frac{\tau_e}{\kappa})} \left[ \frac{2}{\kappa^2\cosh^2   (\frac{\tau_e}{\kappa})}
-\frac{1}{2}(\vec{n}')^2+\frac{\kappa^2}{16}(\vec{n}')^4+\mathcal{O}(\kappa^4)\right]
\end{multline}
We can now evaluate the integral over $\tau_e$. It is clear that the integral over $\tau_e$ peaks around $\tau_e=0$ in the $\kappa \rightarrow 0$ limit. However, we can get a possible contribution from expanding the integrand around $\tau_e=0$.%
\footnote{We thank K. Zarembo for pointing this out to us.} 
Consider the part $G(\tau_e,\sigma)\equiv(e^{-i\varphi}\cos\theta)^{j}$.We expand
\begin{equation}
G(\tau_e,\sigma) = G(0,\sigma) +  \tau_e \frac{\partial G}{\partial \tau_e} \Big|_{(0,\sigma)}  + \frac{1}{2}  \tau_e^2 \frac{\partial^2 G}{\partial \tau_e^2} \Big|_{(0,\sigma)} + \cdots
\end{equation}
The first correction gives zero when integrated over $\tau_e$ since it is an odd function of $\tau_e$. The second correction gives a non-zero contribution instead, but since $\tau_e \sim \kappa$ and because of the Frolov-Tseytlin limit, this correction is of order $\lambda'{}^2$. For this reason we see that no other part of the integrand will pick up a contribution in this way, since they are of higher order in $\lambda'$ and we consider only terms up to order $\lambda'{}^2$.

Using the EOMs for the Landau-Lifshitz sigma-model we compute
\begin{equation}
\frac{\partial^2 G}{\partial \tau_e^2} \Big|_{(0,\sigma)} = \kappa^2 (e^{-i \varphi} \cos \theta  )^j ( K_1 + K_2 )
\end{equation}
with
\begin{equation}
K_1 = - \frac{j(j-1)}{4} \left[ \frac{i \theta'' }{\cos \theta} + \sin \theta ( 2 \tan \theta \theta' \varphi' + i \varphi'{}^2 - \varphi'' ) \right]^2
\end{equation}
\begin{eqnarray}
K_2 &=& - \frac{j}{16} \frac{1}{\cos \theta} \Big\{ 8 i \sin \theta\, \theta'{}^3 \varphi' - 4 \cos \theta\, \sin^2 \theta\, \varphi'{}^4 - 4\cos 3 \theta \,\varphi''{}^2    +4 \theta'^2 [  \sin \theta \, \theta'' - 5 i \cos \theta\, \varphi''  \nn \\ && - 15 \cos \theta \,\sin^2 \theta\, \varphi'{}^2 ]  + \varphi'{}^2 [ ( 5 \sin 3\theta - 19 \sin \theta  ) \theta'' + i ( 7  \cos 3 \theta -3 \cos \theta  ) \varphi'' ]   \nn \\ && + 4 \theta' [ \varphi' (  \sin \theta ( -4 i \cos 2 \theta \, \varphi'{}^2 + ( 11 \cos 2 \theta-1 ) \varphi'' - 6 i \cos \theta\, \theta''  )) - 4 i \sin \theta\,  \varphi''' ]  \nn \\ && 
+ 8 \sin \theta\, \varphi' ( \sin 2\theta \, \varphi''' -2i \theta''' ) + 4\sin \theta ( \theta''''  -6i \theta'' \varphi'') - 4 \cos \theta ( \theta''{}^2 - i \varphi''{}^2 )
\Big\}
\end{eqnarray}
There are three types of integral to perform that we denote as $I_0$, $I_1$ and $I_2$ and they are given by 
\begin{equation} 
I_0= \frac{1}{\kappa} \int_{-\infty}^{+\infty} \frac{d{\tau_e}}{\cosh^{2j+2}  (\frac{\tau_e}{\kappa})} =\frac{2^{2j+1} \left( j! \right)^2}{\left( 2j+1\right)!}
\label{leading}
\end{equation}
\begin{equation}
I_1= \frac{1}{\kappa} \int_{-\infty}^{+\infty} \frac{d{\tau_e}}{\cosh^{2j} (\frac{\tau_e}{\kappa})} =I_0 \frac{2j+1}{2j}
\label{1loop}
\end{equation}
\begin{equation}
I_2= \frac{1}{\kappa^3} \int_{-\infty}^{+\infty} \frac{\tau_e^2 d{\tau_e}}{\cosh^{2j+2}  (\frac{\tau_e}{\kappa})} = I_0 \frac{1}{4} \Psi ( 1 , 1 +j) 
\label{2loop}
\end{equation}
where $\Psi ( 1, x ) = \frac{d^2}{dx^2} \log \Gamma( x)$.
Using this, our final result is 
\begin{eqnarray}
\label{3ptstringfinal2}
C_{123} &=&  \frac{J}{N} \frac{j!}{2^j\sqrt{(2j-1)!} }\int_{0}^{2 \pi} \frac{d\sigma}{2\pi} (e^{-i\varphi}\cos\theta)^{j}  \left[ 1
-\frac{(2j+1)}{4j}\left(\frac{\lambda'}{2}(\vec{n}')^2-\frac{\lambda'^2}{16}(\vec{n}')^4\right) \right. \nn \\[3mm] && \left. +  \frac{\lambda'{}^2}{4} \Psi ( 1 , 1 +j) ( K_1 + K_2 ) +\mathcal{O}(\lambda'^3) \right]
\end{eqnarray}
where we used $\kappa^2 = \lambda'$.


\subsection{Comparison with gauge theory at one loop order}

Having the result \eqref{3ptstringfinal2}, we are ready to make the comparison with the gauge theory result \eqref{C123oneloop}. To this end, it is convenient to compute Eq.~\eqref{C123string} in terms of a different set of coordinates. We do this in this section where we limit ourselves to consider only up to and including the one loop correction.

We write the parametrization of the $3$-sphere using the unitary vector ${\bf{U}}(\sigma, \tau)=e^{i\tau/\kappa} {\bf{u}}(\sigma, \tau)$, where ${\bf u}(\sigma,\tau)=(u_1(\sigma, \tau),u_2(\sigma, \tau),u_3(\sigma, \tau))$.
The limit \eqref{TMlimit} then is
\begin{equation}
\label{limit}
\kappa\rightarrow 0~~,~~\frac{1}{\kappa}\partial_{\tau}{\bf{u}}~~{\rm fixed}, ~~\partial_\sigma {\bf{u}}~~{\rm fixed}
\end{equation}
In this limit the EOMs and Virasoro constraints 
 reduce to $\frac{2\,i}{\kappa}\partial_{\tau}{\bf{u}}=\partial^2_{\sigma}{\bf{u}}+ 2 {\bf u}\left(\partial_{\sigma}\bar{{\bf{u}}}\cdot\partial^{\sigma}{\bf{u}}\right)$ and $\bar{{\bf u}}\cdot \partial_{\sigma}{\bf u}=0$. 
The holographic 3-point function coefficient can thus be computed from
\begin{equation}
C_{123}=  c_j \frac{ \sqrt{\lambda} }{N}\int_{-\infty}^{+\infty} d{\tau_e} \int_{0}^{2 \pi} \frac{d\sigma}{2\pi}\frac{ \left(\bar{u}^1 u_2 \right)^{j}}{\cosh^{2j} (\frac{\tau_e}{\kappa})} \left[ \frac{1}{\kappa^2\cosh^2 (\frac{\tau_e}{\kappa})}-\partial_{\sigma}{ \bar {\bf{u}}} \cdot \partial_{\sigma} { {\bf{u}}}+\mathcal{O}(\kappa^2)\right]
\label{jeq}
\end{equation}
Note that this is the same expression used in \cite{Escobedo:2011xw} if one replaces $\kappa$ with $1/\kappa$ . 
Evaluating the integral over $\tau_e$ as before we obtain
\begin{equation}
C_{123}=  \frac{ J}{N} \, \frac{j!}{\sqrt{(2j-1)!}}\int_{0}^{2 \pi} \frac{d\sigma}{2\pi}\left(\bar{u}^1 u_2 \right)^{j} \left[ 1-\lambda'\frac{2j+1}{2j}\,\partial_{\sigma}{ \bar {\bf{u}}}\cdot \partial_{\sigma} { {\bf{u}}}+\mathcal{O}(\lambda'^2)\right]
\label{fineq}
\end{equation}

Comparing this result with the one for the dual gauge theory, Eq.~\eqref{gaugeC123}, it is evident that the leading terms in both sides perfectly match as already pointed out in \cite{Escobedo:2011xw}. However, it is just as clear that this matching does not extend to the one loop term. 

As pointed out in the Introduction, it is not clear that one should have expected the gauge theory result \eqref{gaugeC123} and the string theory result \eqref{fineq} to match. This is true even for the leading order part corresponding to tree-level on the gauge theory side. However, the fact that the tree-level part does match the string side, certainly raises the hope that also the one-loop part should match. While our analysis seems to conclude that this is not the case, we should point out that there are a number of subtleties in the computations that may not be sufficiently well understood at present in the literature and could therefore possibly affect our results. First of all the possibility of a further contribution to the result \eqref{gaugeC123} coming from the so-called spin-flipped coherent state \cite{Kruczenski:2004kw}, as discussed in Section \ref{sec:spinflip}. Moreover, among the other possible subtleties is the approximation $\CO_1\simeq \bar{\CO}_2$. Indeed, in our computation we have assumed that Eq.~\eqref{vu_app} holds also at two-loop order. Moreover the gauge theory side of the computation, which is based on the prescription of \cite{Okuyama:2004bd,Roiban:2004va,Alday:2005nd}, might still require some explicit tests on the line of the ones performed in \cite{Grossardt:2010xq}. Finally, with our current understanding of the AdS/CFT correspondence, it is not clear whether or not one should expect a matching of the two quantities.
In the case of 3-point functions the comparison between results obtained on the gauge theory and string theory sides has been done only in few cases \cite{Bissi:2011dc, Escobedo:2011xw, Georgiou:2011qk} and it has revealed a very useful instrument for improving our understanding of both the gauge theory and the string theory side.  It would be extremely interesting to push this program forward.

\section*{Acknowledgments}

We thank Kolya Gromov, Romuald Janik, Martin Kruczenski, Amit Sever, Pedro Vieira and Kostya Zarembo for interesting and stimulating discussions. We thank Jan Plefka for useful correspondence. Finally, we thank Charlotte Kristjansen for interesting discussions and for reading the manuscript. TH thanks the Niels Bohr Institute for kind hospitality.

\addcontentsline{toc}{section}{References}


\small

\providecommand{\href}[2]{#2}\begingroup\raggedright\endgroup

\end{document}